\documentclass[12pt]{article}
\usepackage{epsfig,cite,amsmath}

\def\beq{\begin{equation}}
\def\eeq{\end{equation}}
\def\beqar{\begin{eqnarray}}
\def\eeqar{\end{eqnarray}}
\def\barr#1{\begin{array}{#1}}
\def\earr{\end{array}}
\def\bfi{\begin{figure}}
\def\efi{\end{figure}}
\def\btab{\begin{table}}
\def\etab{\end{table}}
\def\bce{\begin{center}}
\def\ece{\end{center}}

\def\text{\textstyle}

\def\al{\alpha}

\def\de{\delta}

\def\si{\sigma}

\def\Ga{\Gamma}
\def\De{\Delta}

\def\refeq#1{\mbox{eq.~(\ref{#1})}}
\def\refeqs#1{\mbox{eqs.~(\ref{#1})}}
\def\reffi#1{\mbox{Fig.~\ref{#1}}}

\def\refta#1{\mbox{Table~\ref{#1}}}

\def\citere#1{\mbox{Ref.~\cite{#1}}}
\def\citeres#1{\mbox{Refs.~\cite{#1}}}

\def\mathswitchr#1{\relax\ifmmode{\mathrm{#1}}\else$\mathrm{#1}$\fi}

\newcommand{\PW}{\mathswitchr W}
\newcommand{\PZ}{\mathswitchr Z}

\newcommand{\PH}{\mathswitchr H}

\newcommand{\Pb}{\mathswitchr b}

\newcommand{\Pt}{\mathswitchr t}

\def\mathswitch#1{\relax\ifmmode#1\else$#1$\fi}

\newcommand{\MW}{\mathswitch {M_\PW}}

\newcommand{\MZ}{\mathswitch {M_\PZ}}
\newcommand{\MH}{\mathswitch {M_\PH}}

\newcommand{\Mb}{\mathswitch {m_\Pb}}
\newcommand{\Mt}{\mathswitch {m_\Pt}}
\newcommand{\scrs}{{}}
\newcommand{\sw}{\mathswitch {s_{\scrs\PW}}}

\newcommand{\cw}{\mathswitch {c_{\scrs\PW}}}
\newcommand{\sweff}{\sin^2 \theta_{\mathrm{eff}}}

\newcommand{\GF}{\mathswitch {G_\mu}}

\newcommand{\mt}{\Mt}

\newcommand{\mb}{\Mb}

\newcommand{\tsf}{\theta\kern-.20em_{\tilde{f}}}
\newcommand{\tsfp}{\theta\kern-.20em_{\tilde{f}\prime}}
\newcommand{\tsq}{\theta\kern-.15em_{\tilde{q}}}

\newcommand{\mf}{m_f}

\newcommand{\lsim}
{\;\raisebox{-.3em}{$\stackrel{\displaystyle <}{\sim}$}\;}

\newcommand{\alps}{\alpha_{\mathrm s}}

\newcommand{\msbar}{$\overline{\rm{MS}}$}

\newcommand{\VL}{\left( \begin{array}{c}}
\newcommand{\VR}{\end{array} \right)}
\newcommand{\ML}{\left( \begin{array}{cc}}
\newcommand{\MLd}{\left( \begin{array}{ccc}}
\newcommand{\MLv}{\left( \begin{array}{cccc}}
\newcommand{\MR}{\end{array} \right)}

\newcommand{\tev}{\,\, \mathrm{TeV}}
\newcommand{\gev}{\,\, \mathrm{GeV}}
\newcommand{\mev}{\,\, \mathrm{MeV}}

\newcommand{\BC}{\begin{center}}
\newcommand{\EC}{\end{center}}
\newcommand{\BE}{\begin{equation}}
\newcommand{\EE}{\end{equation}}
\newcommand{\BEA}{\begin{eqnarray}}
\newcommand{\BEAnn}{\begin{eqnarray*}}
\newcommand{\EEA}{\end{eqnarray}}
\newcommand{\EEAnn}{\end{eqnarray*}}
\newcommand{\non}{\nonumber}
\newcommand{\id}{{\rm 1\kern-.12em
\rule{0.3pt}{1.5ex}\raisebox{0.0ex}{\rule{0.1em}{0.3pt}}}}

\def\draftdate{\relax}
\def\mda{\relax}
\def\mua{\relax}
\def\mla{\relax}
\def\draft{
\def\thtystars{******************************}
\def\sixtystars{\thtystars\thtystars}
\typeout{}
\typeout{\sixtystars**}
\typeout{* Draft mode!
         For final version remove \protect\draft\space in source file
*}
\typeout{\sixtystars**}
\typeout{}
\def\draftdate{\today}
\def\mua{\marginpar[\boldmath\hfil$\uparrow$]%
                   {\boldmath$\uparrow$\hfil}%
                    \typeout{marginpar: $\uparrow$}\ignorespaces}
\def\mda{\marginpar[\boldmath\hfil$\downarrow$]%
                   {\boldmath$\downarrow$\hfil}%
                    \typeout{marginpar: $\downarrow$}\ignorespaces}
\def\mla{\marginpar[\boldmath\hfil$\rightarrow$]%
                   {\boldmath$\leftarrow $\hfil}%
                    \typeout{marginpar:
$\leftrightarrow$}\ignorespaces}
\def\Mua{\marginpar[\boldmath\hfil$\Uparrow$]%
                   {\boldmath$\Uparrow$\hfil}%
                    \typeout{marginpar: $\Uparrow$}\ignorespaces}
\def\Mda{\marginpar[\boldmath\hfil$\Downarrow$]%
                   {\boldmath$\Downarrow$\hfil}%
                    \typeout{marginpar: $\Downarrow$}\ignorespaces}
\def\Mla{\marginpar[\boldmath\hfil$\Rightarrow$]%
                   {\boldmath$\Leftarrow $\hfil}%
                    \typeout{marginpar:
$\Leftrightarrow$}\ignorespaces}
\overfullrule 5pt
\oddsidemargin -15mm
\marginparwidth 29mm
}

\begin{document}
\thispagestyle{empty}

\def\thefootnote{\fnsymbol{footnote}}

\begin{flushright}
DCPT/03/146, DESY 03-184\\
FERMILAB-Pub-03/239-T\\
IPPP/03/73\\
hep-ph/0311148\\
\end{flushright}

\vspace{.2cm}

\begin{center}

{\large\sc {\bf Precise Prediction for the W-Boson Mass}}

\vspace{0.4cm}

{\large\sc {\bf in the Standard Model}}

\vspace{1cm}

{\sc
M.~Awramik$^{1,2}$%
\footnote{email: awramik@ifh.de}%
, M.~Czakon$^{1,3}$%
\footnote{email: czakon@ifh.de}%
, A.~Freitas$^4$%
\footnote{email: afreitas@fnal.gov}%
~and G.~Weiglein$^5$%
\footnote{email: Georg.Weiglein@durham.ac.uk}
}

\vspace*{1cm}

{\sl
$^1$DESY, Platanenallee 6, D-15738 Zeuthen, Germany

\vspace*{0.4cm}

$^2$Institute of Nuclear Physics, Radzikowskiego 152,
      PL-31342 Cracow, Poland

\vspace*{0.4cm}

$^3$Department of Field Theory and Particle Physics,
      Institute of Physics, University of Silesia, Uniwersytecka 4,
      PL-40007 Katowice, Poland

\vspace*{0.4cm}

$^4$Theoretical Physics Division, Fermilab, P.~O.\ Box 500, Batavia, IL
60510, USA

\vspace*{0.4cm}

$^5$Institute for Particle Physics Phenomenology, University of
Durham,\\
Durham DH1~3LE, UK
}

\end{center}

\vspace*{1cm}

\begin{abstract}
The presently most accurate prediction for the W-boson mass in the Standard
Model is obtained by combining the complete two-loop result with the
known higher-order QCD and electroweak corrections. The numerical impact
of the different contributions is analysed in detail. A simple
parametrisation of the full result is presented, which approximates the
full result for $\MW$ to better than 0.5~MeV for $10 \gev \leq \MH \leq
1 \tev$ if the other parameters are varied within their combined $2 \si$
region around their experimental central values. The different sources
of remaining theoretical uncertainties are investigated. Their effect
on the prediction of $\MW$ is estimated to be about 4~MeV for $\MH \lsim
300$~GeV.
\end{abstract}

\def\thefootnote{\arabic{footnote}}
\setcounter{page}{0}
\setcounter{footnote}{0}

\newpage



The relation between the W-boson mass, $\MW$, the Z-boson mass, $\MZ$,
the fine structure constant, $\al$, and the Fermi constant, $\GF$,
\beq
\MW^2 \left(1 - \frac{\MW^2}{\MZ^2}\right) =
\frac{\pi \al}{\sqrt{2} \GF} \left(1 + \De r\right),
\label{eq:delr}
\eeq
is of central importance for precision tests of the electroweak theory.
Accordingly, a lot of effort has been devoted over more than two decades
to accurately predict the quantity $\De r$, which summarises the radiative 
corrections, within the Standard Model (SM) and extensions of it.  

The one-loop result~\cite{sirlin} can be written as
\beq
\De r^{(\al)} = \De \al - \frac{\cw^2}{\sw^2} \De\rho +
\De r_{\mathrm{rem}}(\MH),
\label{eq:delrol}
\eeq
where $\cw^2 = \MW^2/\MZ^2$, $\sw^2 = 1 - \cw^2$. It involves large 
fermionic contributions from the shift in the fine structure
constant due to light fermions, $\De\al \propto \log \mf$,
and from the leading contribution to the $\rho$~parameter, $\De\rho$.
The latter is quadratically dependent on the top-quark mass, $\mt$, as a 
consequence of the large mass splitting in the isospin doublet~\cite{velt}.
The remainder part, $\De r_{\mathrm{rem}}$, contains
in particular the dependence on the Higgs-boson mass, $\MH$.
Higher-order QCD corrections to $\De r$ are known at ${\cal O}(\al
\alps)$~\cite{qcd2} and ${\cal O}(\al\alps^2)$~\cite{qcd3,qcd3light}, as well as
${\cal O}(\al\alps^3)$ for $\Delta\rho$~\cite{qcd4}.

Recently the full electroweak two-loop result for $\De r$ has been
completed. It consists of the fermionic
contribution~\cite{2lferm,2lfermb,2lfermc}, which involves
diagrams with one or two closed fermion loops, and the purely bosonic
two-loop contribution~\cite{2lbos}.

Beyond two-loop order the results for the pure fermion-loop
corrections (i.e.\ contributions containing $n$ fermion loops at
$n$-loop order) are known up to four-loop order~\cite{floops}. They
contain in particular the leading contributions in $\De\al$ and
$\De\rho$. Most recently results for the leading three-loop contributions of 
${\cal O}(\GF^3 \Mt^6)$ and ${\cal O}(\GF^2 \alps \Mt^4)$ have been
obtained for arbitrary values of $\MH$ (by means of expansions around 
$\MH = \mt$ and for $\MH \gg \mt$)~\cite{faisst}, generalising a
previous result which was obtained in the limit $\MH =
0$~\cite{faisstold}.

Eq.~(\ref{eq:delr}) is usually employed for predicting the W-boson
mass,
\beq
\MW^2 = \MZ^2 \left\{\frac{1}{2} + \sqrt{\frac{1}{4} -
        \frac{\pi \al}{\sqrt{2} \GF \MZ^2} \Bigl[1 + \De r
        (\MW, \MZ, \MH, \mt, \dots) \Bigr]}\, \right\} ,
\label{eq:MW}
\eeq
which is done by an iterative procedure since $\De r$ itself depends on
$\MW$. Comparison of the prediction for $\MW$ within the SM with the
experimental value allows to obtain indirect constraints on the
Higgs-boson mass. These constraints are affected both by the experimental 
error of $\MW$ and by the uncertainty of the theory prediction. 
The current experimental error of the W-boson mass is
$\de\MW^{\rm exp} = 34$~MeV~\cite{datasummer2003}. The
accuracy in the measurement of the W-boson mass is expected to improve 
to about $\de\MW^{\rm exp, Tev/LHC} = 15$~MeV~\cite{mwlhc} from the
measurements at RunII of the Tevatron and the LHC, and to about
$\de\MW^{\rm exp, LC} = 7$~MeV at a future
Linear Collider (LC) running at the WW~threshold~\cite{mtmwlc}.
The uncertainty of the theory prediction
is caused by the experimental errors of the input parameters,
e.g.\ $\mt$, and by the uncertainty from unknown higher-order
corrections. In the global SM fit to all data~\cite{quast03}
the highest sensitivity to $\MH$ arises from the predictions for $\MW$ and 
the effective weak mixing angle at the Z-boson resonance, $\sweff$.

In the present paper we combine the various pieces being relevant for the
prediction of $\MW$ into a common result and analyse the numerical
impact of the different contributions. Since in particular the
electroweak two-loop result is very lengthy and involves numerical
integrations of two-loop scalar integrals, it is not possible to present
the full result in a compact analytic form. We therefore provide a simple
parametrisation of the full result which is easy to implement and
should be accurate enough for practical applications. We discuss the
sources of the remaining theoretical uncertainties and obtain an
estimate for the uncertainty from unknown higher-order corrections. 

\bigskip


We incorporate the following contributions into the result for $\De r$,
\beq
\De r = \De r^{(\al)} + \De r^{(\al\alps)} + \De r^{(\al\alps^2)} +
\De r^{(\al\alps^3\Mt^2)} + 
\De r^{(\al^2)}_{\rm ferm} + \De r^{(\al^2)}_{\rm bos} +
\De r^{(\GF^2 \alps \Mt^4)} + \De r^{(\GF^3 \Mt^6)} ,
\label{eq:delrcontribs}
\eeq
where $\De r^{(\al)}$ is the one-loop result, \refeq{eq:delrol}, $\De
r^{(\al\alps)}$, $\De r^{(\al\alps^2)}$ and $\De r^{(\al\alps^3\mt^2)}$ 
are the two-loop~\cite{qcd2}, three-loop~\cite{qcd3,qcd3light} and
approximate four-loop~\cite{qcd4} QCD corrections, and 
$\De r^{(\al^2)}_{\rm ferm}$~\cite{2lferm,2lfermb,2lfermc} and 
$\De r^{(\al^2)}_{\rm bos}$~\cite{2lbos} are the fermionic and purely 
bosonic electroweak two-loop corrections, respectively. 
The contributions $\De r^{(\GF^2 \alps \Mt^4)}$ and 
$\De r^{(\GF^3 \Mt^6)}$ have been obtained from the leading three-loop
contributions to $\De\rho$ given in \citere{faisst}. 

We have not included the pure fermion-loop contributions at three-loop
and four-loop order obtained in \citere{floops}, because their
contribution turned out to be small as a consequence of accidental
numerical cancellations, with a net effect of only about 1~MeV in $\MW$
(using the real-pole definition of the gauge-boson masses). Since the
result given in \citere{floops} contains the leading contributions
involving powers of $\De\al$ and $\De\rho$ beyond two-loop order, we do
not make use of resummations of $\De\al$ and $\De\rho$ as it was often
done in the literature in the past (see e.g.\ \citeres{resum}).
Accordingly, the quantity $\De r$ appears in \refeq{eq:MW} in fully
expanded form. This means, for instance, that we do not include the 
${\cal O}(\al^3)$ term $3 (\De\al)^2 \De r^{(\al)}_{\rm bos}$, which 
can be inferred from the electric charge renormalisation. It affects the 
prediction for $\MW$ by about $1.5$~MeV. This shift is however expected
to partially cancel with the corresponding contributions proportional to
$(\De\al) (\De\rho) \De r^{(\al)}_{\rm bos}$ and
$(\De\rho)^2 \De r^{(\al)}_{\rm bos}$ in an analogous way as for the
pure fermion-loop contributions.

In \refta{tab:delrcontribs} the numerical values of the different
contributions to $\De r$ are given for $\MW =
80.426$~GeV~\cite{datasummer2003}. The other input parameters that we
use in this paper are~\cite{datasummer2003}\footnote{The value for
\GF\ has been updated to the 2014 value \cite{pdg14}.}
\beqar
&& \mt = 174.3 \gev, \quad \mb = 4.7 \gev, \quad \MZ = 91.1875 \gev, \quad
\Ga_{\PZ} = 2.4952 \gev, \non \\
&& \al^{-1} = 137.03599976, \quad
\De\al = 0.05907, \quad \alps(\MZ) = 0.119, \non \\
&& \GF = 1.166379 \times 10^{-5} \gev^{-2} , 
\label{eq:inputs}
\eeqar
where $\De\al \equiv \De\al_{\rm lept} + \De\al^{(5)}_{\rm had}$, and
$\De\al_{\rm lept} = 0.0314977$~\cite{delalplept}. For
$\De\al^{(5)}_{\rm had}$ we use the value given in \citere{fjeg},
$\De\al^{(5)}_{\rm had} = 0.027572 \pm 0.000359$. The total width of the 
Z~boson, $\Ga_{\PZ}$, appears as an input parameter since the
experimental value of $\MZ$ in \refeq{eq:inputs}, corresponding to a
Breit--Wigner parametrisation with running width, needs to be
transformed in our calculation into the mass parameter defined according
to the real part of the complex pole, which corresponds to a
Breit--Wigner parametrisation with a constant decay width, see
\citere{2lfermb}. It is understood that $\MW$ in this paper always
refers to the conventional definition according to a Breit--Wigner
parametrisation with running width. The change of parametrisations is
achieved with the one loop QCD corrected value of the W-boson width as
described in \citere{2lfermb}.

\btab[tp]
$$
\begin{array}{|c||c|c|c|c|c|c|c|c|} \hline
\MH /{\rm GeV} &
\De r^{(\al)} & \De r^{(\al\alps)} & \De r^{(\al\alps^2)} &
\De r^{(\al\alps^3\mt^2)} &
\De r^{(\al^2)}_{\rm ferm} & \De r^{(\al^2)}_{\rm bos}  &
\De r^{(\GF^2 \alps \Mt^4)} & \De r^{(\GF^3 \Mt^6)}
\\ \hline
100  & 283.41 & 35.89 & 7.23 & 1.27 & 28.56 & 0.64  & -1.27 & -0.16 \\
200  & 307.35 & 35.89 & 7.23 & 1.27 & 30.02 & 0.35  & -2.11 & -0.09 \\
300  & 323.27 & 35.89 & 7.23 & 1.27 & 31.10 & 0.23  & -2.77 & -0.03 \\
600  & 353.01 & 35.89 & 7.23 & 1.27 & 32.68 & 0.05  & -4.10 & -0.09 \\
1000 & 376.27 & 35.89 & 7.23 & 1.27 & 32.36 & -0.41 & -5.04 & -1.04 \\ \hline
\end{array}
$$
\caption{The numerical values ($\times 10^4$) of the different 
contributions to $\De r$ specified in \refeq{tab:delrcontribs} are given 
for different values of $\MH$ and 
$\MW = 80.426$~GeV (the W and Z masses have been transformed so as to
correspond to the real part of the complex pole). The other input
parameters are listed in \refeq{eq:inputs}.
\label{tab:delrcontribs}}
\etab

\refta{tab:delrcontribs} shows that the two-loop QCD correction, 
$\De r^{(\al\alps)}$, and the fermionic electroweak two-loop correction,
$\De r^{(\al^2)}_{\rm ferm}$ are of similar size. They both amount to
about 10\% of the one-loop contribution, $\De r^{(\al)}$, 
entering with the same sign.
The most important correction beyond these contributions is the
three-loop QCD correction, $\De r^{(\al\alps^2)}$, which leads to a
shift in $\MW$ of about $-11$~MeV. For large values of $\MH$ also the 
contribution $\De r^{(\GF^2 \alps \Mt^4)}$ becomes sizable. The purely 
bosonic two-loop contribution, $\De r^{(\al^2)}_{\rm bos}$, and the leading
electroweak three-loop correction, $\De r^{(\GF^3 \Mt^6)}$, 
and leading QCD four-loop correction, $\De r^{(\al\alps^3\mt^2)}$,
give rise to
shifts in $\MW$ which are significantly smaller than the experimental
error envisaged for a future Linear Collider, 
$\de\MW^{\rm exp, LC} = 7$~MeV~\cite{mtmwlc}. 

Since $\De r$ is evaluated in \refta{tab:delrcontribs} for a fixed value 
of $\MW$, the contributions $\De r^{(\al\alps)}$ and $\De r^{(\al\alps^2)}$
are $\MH$-independent. In the iterative procedure for evaluating $\MW$
according to \refeq{eq:MW}, on the other hand, also these contributions
become $\MH$-dependent through the $\MH$-dependence of the inserted
$\MW$ value.

The result for $\MW$ based on \refeqs{eq:MW}, (\ref{eq:delrcontribs})
can be approximated by the following simple parametrisation (see
\citere{mwparamold} for an earlier parametrisation of $\MW$),
\beqar
\MW &=& \MW^0 - c_1 \, \mathrm{dH} - c_2 \, \mathrm{dH}^2 
       + c_3 \, \mathrm{dH}^4 + c_4 (\mathrm{dh} - 1)
       - c_5 \, \mathrm{d}\al + c_6 \, \mathrm{dt} 
       - c_7 \, \mathrm{dt}^2 \non \\
&& {}  - c_8 \, \mathrm{dH} \, \mathrm{dt} 
       + c_9 \, \mathrm{dh} \, \mathrm{dt} - c_{10} \, \mathrm{d}\alps
       + c_{11} \, \mathrm{dZ} ,
\label{eq:fitformula}
\eeqar
where
\beqar
\mathrm{dH} = \ln\left(\frac{\MH}{100 \gev}\right), &&
\mathrm{dh} = \left(\frac{\MH}{100 \gev}\right)^2, \quad
\mathrm{dt} = \left(\frac{\mt}{174.3 \gev}\right)^2 - 1, \non \\
\mathrm{dZ} = \frac{\MZ}{91.1875 \gev} -1, &&
\mathrm{d}\al = \frac{\De\al}{0.05907} - 1, \quad
\mathrm{d}\alps = \frac{\alps(\MZ)}{0.119} - 1 ,
\label{eq:pardef}
\eeqar
and the coefficients $\MW^0, c_1, \ldots, c_{11}$ take the following values
\beqar
\MW^0 = 80.3779 \gev, & c_1 = 0.05427 \gev, & c_2 = 0.008931 \gev , \non \\
c_3 = 0.0000882 \gev, & c_4 = 0.000161 \gev, & c_5 = 1.070 \gev , \non \\
c_6 = 0.5237 \gev, & c_7 = 0.0679 \gev, & c_8 = 0.00179 \gev , \non \\
c_9 = 0.0000664 \gev, & c_{10} = 0.0795 \gev, & c_{11} = 114.9 \gev .
\label{eq:fitparams}
\eeqar
The parametrisation given in
\refeqs{eq:fitformula}--(\ref{eq:fitparams}) approximates the full
result for $\MW$ to better than 0.5~MeV over the whole range of 
$10 \gev \leq \MH \leq 1 \tev$ if all other experimental input values
vary within their combined $2 \si$ region around their central values 
given in \refeq{eq:pardef}.

\btab[tp]
$$
\begin{array}{|c||c|c|} \hline
 & \de\MW(\mbox{full result})/\mev & 
   \de\MW(\refeqs{eq:fitformula}\mbox{--}(\ref{eq:fitparams}))/\mev
\\ \hline
\de\MH = 100 \gev & -41.3 & -41.4 \\
\de\mt = 5.1 \gev & 31.0 & 31.0 \\
\de\MZ = 2.1 \mev & 2.6 & 2.6 \\
\de\left(\De\al^{(5)}_{\rm had}\right) = 0.00036 & -6.5 & -6.5 \\
\de\alps(\MZ) = 0.0027 & -1.7 & -1.7 \\
\hline
\end{array}
$$
\caption{Shifts in $\MW$ caused by varying $\MH$ by 100~GeV and the
other input parameters by $1 \si$ around their experimental central
values~\cite{datasummer2003}. The first column shows the full result for
$\MW$, while the second column is based on the simple parametrisation of 
\refeqs{eq:fitformula}--(\ref{eq:fitparams}).
The shifts $\de\MW$ are relative to the value 
$\MW = 80.3799$~GeV which is the result for $\MH = 100$~GeV and the
central values of the other input parameters as specified in
\refeq{eq:inputs}.
\label{tab:paramvars}}
\etab

In \refta{tab:paramvars} the full result for $\MW$ and the
parametrisation of \refeqs{eq:fitformula}--(\ref{eq:fitparams}) are
compared with each other. The table shows the shifts in $\MW$ (relative
to the value $\MW = 80.3799$~GeV, which is the result for $\MH =
100$~GeV and the central values of the other input parameters as specified in
\refeq{eq:inputs}) induced by varying $\MH$ by 100~GeV and the other
input parameters by $1 \si$ around their experimental central
values~\cite{datasummer2003}. In the example of \refta{tab:paramvars},
where only one parameter has been varied in each row and all others have
been kept at their central values, the maximum deviation between the
full result for $\MW$ and the parametrisation of
\refeqs{eq:fitformula}--(\ref{eq:fitparams}) is below 0.1~MeV.

The parametrisation of \refeqs{eq:fitformula}--(\ref{eq:fitparams})
yields a good approximation of the full result for $\MW$ even for values
of $\MH$ much smaller than the experimental 95\% C.L.\ lower bound on the 
Higgs-boson mass, $\MH = 114.4$~GeV~\cite{mhlimit}. If one restricts to 
the region $\MH > 100$~GeV, a slight readjustment of the coefficients in 
\refeq{eq:fitparams} yields an even more accurate parametrisation of the
full result. If \refeqs{eq:fitformula}, (\ref{eq:pardef}) are used
together with the following values of the coefficients,
\beqar
\MW^0 = 80.3779 \gev, & c_1 = 0.05263 \gev, & c_2 = 0.010239 \gev , \non \\
c_3 = 0.000954 \gev, & c_4 = -0.000054 \gev, & c_5 = 1.077 \gev , \non \\
c_6 = 0.5252 \gev, & c_7 = 0.0700 \gev, & c_8 = 0.004102 \gev , \non \\
c_9 = 0.000111 \gev, & c_{10} = 0.0774 \gev, & c_{11} = 115.0 \gev ,
\label{eq:fitparams2}
\eeqar
the full result for $\MW$ is approximated to better than 0.25~MeV over
the range of $100 \gev \leq \MH \leq 1 \tev$ if all other experimental
input values vary within their combined $2 \si$ region around their 
central values given in \refeq{eq:pardef}.

{}From \refta{tab:paramvars} one can read off the parametric theoretical
uncertainties in the prediction for $\MW$ being caused by the
experimental errors of the input parameters. The dominant parametric 
uncertainty at present (besides the dependence on $\MH$) 
is induced by the experimental error of the
top-quark mass. It is almost as large as the current experimental error
of the W-boson mass, $\de\MW^{\rm exp} = 34$~MeV~\cite{datasummer2003}.
The uncertainty caused by the experimental error of $\mt$ will remain
the dominant source of theoretical uncertainty in the prediction for
$\MW$ even at the LHC, where the error on $\mt$ will be reduced to
$\de\mt = 1$--2~GeV~\cite{mtlhc}. A further improvement of the
parametric uncertainty of $\MW$ will require the precise measurement of 
$\mt$ at a future Linear Collider~\cite{delmt}, where an accuracy
of about $\de\mt = 0.1$~GeV will be achievable~\cite{mtmwlc}. 

We now turn to the second source of theoretical uncertainties in the
prediction for $\MW$, namely the uncertainties from unknown higher-order 
corrections. Different approaches have been used in the literature for
estimating the possible size of uncalculated higher-order 
corrections~\cite{bluebandmethod,mwest,snowmass,2lfermb}. 
The ``traditional Blue Band method'' is based on the fact that the
results of calculations employing different renormalisation schemes or
different prescriptions for including non-leading contributions in
resummed or expanded form differ from each other by higher-order
corrections. The deviations between the results of different codes in
which the same corrections have been organised in a somewhat different
way are used in this method as a measure for the size of unknown
higher-order corrections~\cite{bluebandmethod}. In applying this method
it is not easy to quantify how big the variety of different ``options''
and different codes should be in order to obtain a reasonable estimate
of the higher-order uncertainties. As the method cannot account for
genuine effects of irreducible higher-order corrections, it may lead to
an underestimate of the theoretical uncertainties if at an uncalculated
order a new source of potentially large corrections appears, e.g.\ a
certain enhancement factor.

In \citere{snowmass} a different prescription has been proposed, in
which for each type of unknown corrections the relevant enhancement
factors are identified and the remaining coefficient arising from the
actual loop integrals is set to unity. In \citere{2lfermb} higher-order
QCD corrections have been estimated in two different ways, from the
renormalisation scale dependence (in particular taking into account the
effect of switching from the on-shell to the \msbar\ definition of the
top-quark mass) and from assuming that, for instance, the ratio of the 
${\cal O}(\al^2\alps)$ and ${\cal O}(\al^2)$ corrections is of the same
size as the ratio of the ${\cal O}(\al \alps)$ and ${\cal O}(\al)$
corrections.

Several of the corrections whose possible size had been estimated in
\citeres{mwest,snowmass,2lfermb} have meanwhile been
calculated~\cite{2lbos,faisst}, and it turned out that the estimates
agree reasonably well with the actual size of the corrections. This adds
confidence to applying the same kind of methods also for an estimate of
the remaining higher-order uncertainties. 

There are two main sources of remaining uncertainties in the prediction for
$\MW$ from unknown higher-order corrections:
\begin{itemize}
\item
The corrections at ${\cal O}(\al^2\alps)$ beyond the known contribution
of ${\cal O}(\GF^2 \alps \Mt^4)$:\\
The numerical effect of the ${\cal O}(\GF^2 \alps \Mt^4)$ correction was 
found to be up to 5~MeV in $\MW$ for a light Higgs-boson mass, $\MH
\lsim 300$~GeV~\cite{faisst}. This contribution represents the leading term
in an expansion for asymptotically large values of $\mt$. In the
calculation of the electroweak two-loop corrections it was found that
the formally next-to-leading order term of ${\cal O}(\GF^2\mt^2\MZ^2)$
has approximately the same numerical effect as the formally leading term
of ${\cal O}(\GF^2\mt^4)$~\cite{dgs}. It can therefore be expected that
also the formally next-to-leading order term of 
${\cal O}(\GF^2\alps\mt^2\MZ^2)$ may be of similar size as the leading 
${\cal O}(\GF^2 \alps \Mt^4)$ term. We therefore assign an uncertainty
of about 3 MeV to the remaining theoretical uncertainties at ${\cal
O}(\al^2\alps)$ (for $\MH \lsim 300$~GeV).

\item
The unknown electroweak three-loop corrections:\\
The numerical effect of the ${\cal O}(\GF^3 \Mt^6)$ contribution was
found to be small~\cite{faisst}, shifting $\MW$ by less than 0.3~MeV for
$\MH \lsim 300$~GeV. This shift is significantly smaller than the estimate in 
\citere{snowmass}. The pure fermion-loop corrections at
three-loop order were found in \citere{floops} to shift $\MW$ by about 1~MeV,
which however involved an accidental numerical cancellation. It thus
doesn't seem to be justified to assume that all other
electroweak three-loop corrections are completely negligible. In
\citere{2lfermb} it was pointed out that reparametrising the W-boson
width, which enters the prediction for $\MW$ at the two-loop level, by
$\GF$ instead of $\al$ shifts the prediction for $\MW$ by about 1~MeV,
which is formally an effect of ${\cal O}(\al^3)$. In order to take into
account uncertainties of this kind (see also the discussion below 
\refeq{eq:delrcontribs}) we assign an uncertainty of 
2~MeV to the unknown corrections at ${\cal O}(\al^3)$.


\end{itemize}
Adding the above estimates for the different kinds of unknown
higher-order corrections in quadrature, we find as estimate of the
remaining theoretical uncertainties from unknown higher-order
corrections
\beq
\de\MW^{\rm theo} \approx 4 \mev.
\label{eq:theounc}
\eeq
This estimate holds for a relatively light Higgs boson, $\MH \lsim
300$~GeV. For a heavy Higgs boson, i.e.\ $\MH$ close to the TeV scale, the 
remaining theoretical uncertainty is significantly larger.

\begin{figure}[t]
\bce
\includegraphics[width=14cm]{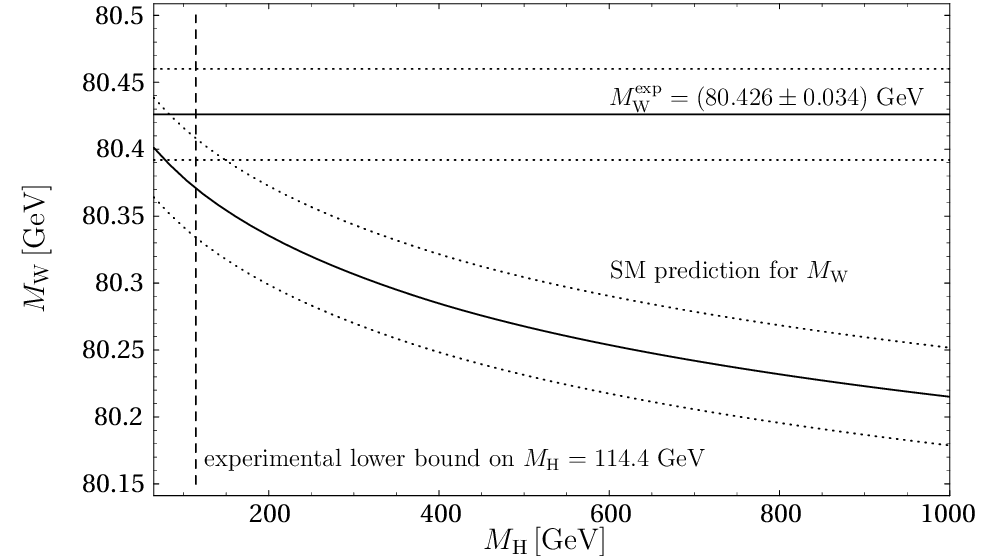}
\ece
\caption{Prediction for $\MW$ in the SM as a function of $\MH$ for $\mt
= 174.3 \pm 5.1$~GeV. The current experimental value,
$\MW^{\mathrm{exp}} = 80.426 \pm 0.034$~GeV~\cite{datasummer2003}, and
the experimental 95\% C.L.\ lower bound on the Higgs-boson mass,
$\MH = 114.4$~GeV~\cite{mhlimit}, are also indicated.
}
\label{fig:MWpred}
\end{figure}

\reffi{fig:MWpred} shows the updated comparison between the theory
prediction for $\MW$ within the SM and the experimental value at the time of the
original publication in 2003. It includes the theory prediction based on 
\refeqs{eq:MW}, (\ref{eq:delrcontribs}), but excluding the ${\cal
O}(\al\alps^3)$ term, and the experimental data from
Ref.~\cite{datasummer2003}.
For the theoretical uncertainty
the estimate of \refeq{eq:theounc} and the parametric uncertainties
corresponding to $1 \si$ variations of the input parameters (see
\refta{tab:paramvars}) have been used. As discussed above, at present
the theoretical uncertainty is dominated by the effect of the
experimental error of the top-quark mass.

Fig.~\ref{fig:MWpred} confirms the well-known preference for a light
Higgs-boson mass within the SM. If the 95\% exclusion bound from the
direct search for the SM Higgs is taken into account~\cite{mhlimit}, 
the $1 \si$ bands corresponding to the theory prediction and the
experimental result for $\MW$ show only a marginal overlap.


\bigskip


In summary, we have presented the currently most accurate prediction
for $\MW$ in the Standard Model. We have discussed the relative
importance of the complete one-loop and two-loop contributions as well
as the known corrections beyond two-loop order. We have summarised the
present status of the theoretical uncertainties of $\MW$ from the
experimental errors of the input parameters, and we have obtained an
estimate for the remaining theoretical uncertainties from unknown
higher-order corrections. In the region of Higgs-mass values preferred
by the electroweak precision data, $\MH \lsim 300$~GeV, the uncertainty
from unknown higher-order corrections amounts to about 4~MeV. This is
much smaller than the present experimental error of $\MW$ and even below
the envisaged future experimental error at the next generation of
colliders. Having reached this level of theoretical precision of $\MW$
is important, however, for the precision test of the electroweak theory,
in particular in view of the fact that $\MW$ can be used as an input for
calculating the effective weak mixing angle at the Z~resonance,
$\sweff$.

We have furthermore presented a simple parametrisation of the full
result containing all relevant corrections, which should be sufficiently
accurate for practical applications. It approximates the full
result for $\MW$ to better than 0.5~MeV over the whole range of 
$10 \gev \leq \MH \leq 1 \tev$ if all other experimental input values
vary within their combined $2 \si$ region around their experimental
central values. In view of the experimental exclusion bound on the
Higgs-boson mass of $\MH > 114.4$~GeV it will normally be sufficient
to restrict to the smaller range of $100 \gev \leq \MH \leq 1 \tev$.
For this case we provide a simple parametrisation which approximates the
full result for $\MW$ even within 0.25~MeV.


\subsection*{Acknowledgements}

This work was supported in part by TMR, EC-Contracts
No.\ HPRN-CT-2002-00311 (EURIDICE) and 
HPRN-CT-2000-00149 (Physics at Colliders),
and the KBN Grants 2P03B01025 and 5P03B09320.
M.A., M.C.\ and A.F.\ would like to thank the Institute for
Particle Physics Phenomenology of the University of
Durham for warm hospitality during the period when
part of this work was completed.




\begin{thebibliography}{00}  

\bibitem{sirlin}
A.~Sirlin,
Phys.\ Rev.\ D {\bf 22} (1980) 971;\\
W.~J.~Marciano and A.~Sirlin,
Phys.\ Rev.\ D {\bf 22} (1980) 2695
[Erratum-ibid.\ D {\bf 31} (1985) 213].

\bibitem{velt}
M.~J.~Veltman,
Nucl.\ Phys.\ B {\bf 123} (1977) 89.

\bibitem{qcd2}
A.~Djouadi and C.~Verzegnassi,
Phys.\ Lett.\ B {\bf 195} (1987) 265;\\
A.~Djouadi,
Nuovo Cim.\ A {\bf 100} (1988) 357;\\
B.~A.~Kniehl,
Nucl.\ Phys.\ B {\bf 347} (1990) 86;\\
F.~Halzen and B.~A.~Kniehl,
Nucl.\ Phys.\ B {\bf 353} (1991) 567;\\
B.~A.~Kniehl and A.~Sirlin,
Nucl.\ Phys.\ B {\bf 371} (1992) 141;\\
B.~A.~Kniehl and A.~Sirlin,
Phys.\ Rev.\ D {\bf 47} (1993) 883;\\
A.~Djouadi and P.~Gambino,
Phys.\ Rev.\ D {\bf 49} (1994) 3499
[Erratum-ibid.\ D {\bf 53} (1994) 4111]
[arXiv:hep-ph/9309298].

\bibitem{qcd3}
L.~Avdeev, J.~Fleischer, S.~Mikhailov and O.~Tarasov,
Phys.\ Lett.\ B {\bf 336} (1994) 560
[Erratum-ibid.\ B {\bf 349} (1994) 597]
[arXiv:hep-ph/9406363];\\
K.~G.~Chetyrkin, J.~H.~K\"uhn and M.~Steinhauser,
Phys.\ Lett.\ B {\bf 351} (1995) 331
[arXiv:hep-ph/9502291];\\
K.~G.~Chetyrkin, J.~H.~K\"uhn and M.~Steinhauser,
Phys.\ Rev.\ Lett.\  {\bf 75} (1995) 3394
[arXiv:hep-ph/9504413].

\bibitem{qcd3light}
K.~G.~Chetyrkin, J.~H.~K\"uhn and M.~Steinhauser,
Nucl.\ Phys.\ B {\bf 482} (1996) 213
[arXiv:hep-ph/9606230].

\bibitem{qcd4}
Y.~Schr\"oder and M.~Steinhauser,
Phys.\ Lett.\ B {\bf 622} (2005) 124
[arXiv:hep-ph/0504055];\\
K.~G.~Chetyrkin, M.~Faisst, J.~H.~K\"uhn, P.~Maierhoefer and C.~Sturm,
  Phys.\ Rev.\ Lett.\  {\bf 97} (2006) 102003
  [arXiv:hep-ph/0605201];\\
  R.~Boughezal and M.~Czakon,
  Nucl.\ Phys.\  B {\bf 755} (2006) 221
  [arXiv:hep-ph/0606232].%

\bibitem{2lferm} 
A.~Freitas, W.~Hollik, W.~Walter and G.~Weiglein,
Phys.\ Lett.\ B {\bf 495} (2000) 338
[Erratum-ibid.\ B {\bf 570} (2003) 260]
[arXiv:hep-ph/0007091].

\bibitem{2lfermb}
A.~Freitas, W.~Hollik, W.~Walter and G.~Weiglein,
Nucl.\ Phys.\ B {\bf 632} (2002) 189
[Erratum-ibid.\ B {\bf 666} (2003) 305]
[arXiv:hep-ph/0202131].

\bibitem{2lfermc}
M.~Awramik and M.~Czakon,
Phys.\ Lett.\ B {\bf 568} (2003) 48
[arXiv:hep-ph/0305248].

\bibitem{2lbos}
M.~Awramik and M.~Czakon,
Phys.\ Rev.\ Lett.\  {\bf 89} (2002) 241801
[arXiv:hep-ph/0208113]; 
{\it see also}
Nucl.\ Phys.\ Proc.\ Suppl.\  {\bf 116} (2003) 238
[arXiv:hep-ph/0211041].\\
A.~Onishchenko and O.~Veretin,
Phys.\ Lett.\ B {\bf 551} (2003) 111
[arXiv:hep-ph/0209010];\\
M.~Awramik, M.~Czakon, A.~Onishchenko and O.~Veretin,
Phys.\ Rev.\ D {\bf 68} (2003) 053004
[arXiv:hep-ph/0209084].

\bibitem{floops}
G.~Weiglein,
Acta Phys.\ Polon.\ B {\bf 29} (1998) 2735
[hep-ph/9807222];\\
A.~Stremplat, Diploma thesis (Univ.\ of Karlsruhe, 1998).

\bibitem{faisst}
M.~Faisst, J.~H.~K\"uhn, T.~Seidensticker and O.~Veretin,
Nucl.\ Phys.\ B {\bf 665} (2003) 649
[arXiv:hep-ph/0302275].

\bibitem{faisstold}
J.~J.~van der Bij, K.~G.~Chetyrkin, M.~Faisst, G.~Jikia and
T.~Seidensticker,
Phys.\ Lett.\ B {\bf 498} (2001) 156
[arXiv:hep-ph/0011373].

\bibitem{datasummer2003}
P.~Wells, talk presented at HEP2003 Europhysics Conference, Aachen, July
2003, to appear in the proceedings.

\bibitem{mwlhc}
S.~Haywood {\it et al.},
arXiv:hep-ph/0003275,
in: Standard Model Physics (and more) at the LHC,
eds.\ G.~Altarelli and M.~Mangano,
CERN, Geneva, 1999 [CERN-2000-004].

\bibitem{mtmwlc}
J.~A.~Aguilar-Saavedra {\it et al.}  [ECFA/DESY LC Physics Working Group
                  Collaboration],
arXiv:hep-ph/0106315;\\
T.~Abe {\it et al.}  [American Linear Collider Working Group
Collaboration],
in {\it Proc. of the APS/DPF/DPB Summer Study on the Future of Particle
Physics (Snowmass 2001) } ed. N.~Graf,
arXiv:hep-ex/0106055;\\
K.~Abe {\it et al.}  [ACFA Linear Collider Working Group Collaboration],
arXiv:hep-ph/0109166;
see: {\tt lcdev.kek.jp/RMdraft/} .

\bibitem{quast03}
G.~Quast, talk presented at HEP2003 Europhysics Conference, Aachen, July
2003, to appear in the proceedings.

\bibitem{resum}
W.~J.~Marciano,
Phys.\ Rev.\ D {\bf 20} (1979) 274;\\
A.~Sirlin,
Phys.\ Rev.\ D {\bf 29} (1984) 89;\\
M.~Consoli, W.~Hollik and F.~Jegerlehner,
Phys.\ Lett.\ B {\bf 227} (1989) 167.

\bibitem{pdg14}
  K.~A.~Olive {\it et al.}  [Particle Data Group Collaboration],
  Chin.\ Phys.\ C {\bf 38} (2014) 090001.

\bibitem{delalplept}
M.~Steinhauser,
Phys.\ Lett.\ B {\bf 429} (1998) 158
[arXiv:hep-ph/9803313].

\bibitem{fjeg}
F.~Jegerlehner,
J.\ Phys.\ G {\bf 29} (2003) 101
[arXiv:hep-ph/0104304].

\bibitem{mwparamold}
G.~Degrassi, P.~Gambino, M.~Passera and A.~Sirlin,
Phys.\ Lett.\ B {\bf 418} (1998) 209
[arXiv:hep-ph/9708311].

\bibitem{mhlimit}
[The LEP working group for Higgs boson searches],
Phys.\ Lett.\ B {\bf 565} (2003) 61
[arXiv:hep-ex/0306033].

\bibitem{mtlhc}
M.~Beneke {\it et al.},
arXiv:hep-ph/0003033,
in: Standard Model Physics (and more) at the LHC,
eds.\ G.~Altarelli and M.~Mangano,
CERN, Geneva, 1999 [CERN-2000-004].

\bibitem{delmt}
S.~Heinemeyer, S.~Kraml, W.~Porod and G.~Weiglein,
JHEP {\bf 0309} (2003) 075
[arXiv:hep-ph/0306181].

\bibitem{bluebandmethod}
D.~Y.~Bardin {\it et al.},
arXiv:hep-ph/9709229;\\
D.~Y.~Bardin, M.~Grunewald and G.~Passarino,
arXiv:hep-ph/9902452.

\bibitem{mwest}
P.~Gambino,
arXiv:hep-ph/9812332;\\
A.~Freitas, S.~Heinemeyer, W.~Hollik, W.~Walter and G.~Weiglein,
Nucl.\ Phys.\ Proc.\ Suppl.\  {\bf 89} (2000) 82
[arXiv:hep-ph/0007129];\\
A.~Ferroglia, G.~Ossola and A.~Sirlin,
Phys.\ Lett.\ B {\bf 507} (2001) 147
[arXiv:hep-ph/0103001].

\bibitem{snowmass}
U.~Baur {\it et al.},
hep-ph/0202001,
in {\it Proc. of the APS/DPF/DPB Summer Study on the Future of Particle
Physics (Snowmass 2001) } eds. R.~Davidson and C.~Quigg.

\bibitem{dgs}
G.~Degrassi, P.~Gambino and A.~Sirlin,
Phys.\ Lett.\ B {\bf 394} (1997) 188
[arXiv:hep-ph/9611363].



\end{thebibliography}
\end{document}